\documentclass[useAMS,usenatbib,usegraphicx]{mn2e}
\usepackage{times}

\newif\ifAMStwofonts
\AMStwofontstrue

\newcommand{\simlt}{\lower.5ex\hbox{$\; \buildrel < \over \sim \;$}}
\newcommand{\simgt}{\lower.5ex\hbox{$\; \buildrel > \over \sim \;$}}
\newcommand{\be}{\begin{equation}}
\newcommand{\ba}{\begin{eqnarray}}
\newcommand{\ee}{\end{equation}}
\newcommand{\ea}{\end{eqnarray}}
 
\newcommand{\Msun}{\mbox{$\rm M_{\odot}$}}

\newcommand{\Lsun}{\mbox{$\>{\rm L_{\odot}}$}}

\newcommand{\etal}{{et~al.~}}

\title[GOODS on the formation of massive galaxies]{On the formation of
massive galaxies: A simultaneous study of number density, size and
intrinsic colour evolution in GOODS}
\author[I. Ferreras \etal]
{Ignacio Ferreras$^{1}$\thanks{E-mail: ferreras@star.ucl.ac.uk},
Thorsten Lisker$^2$, Anna Pasquali$^3$, Sadegh Khochfar$^4$, Sugata Kaviraj$^{1,5}$\\
$^1$ Mullard Space Science Laboratory, Unversity College London, 
Holmbury St Mary, Dorking, Surrey RH5 6NT\\
$^2$ Astronomisches Rechen-Institut, Zentrum f\"ur Astronomie, Universit\"at 
Heidelberg, M\"onchhofstr. 12-14, D-69120 Heidelberg, Germany\\
$^3$ Max-Planck-Institut f\"ur Astronomie, Koenigstuhl 17, D-69117
Heidelberg, Germany\\ 
$^4$ Max-Planck-Institut f\"ur Extraterrestrische Physik, 
Giessenbachstrasse, D-85748 Garching, Germany\\
$^5$ Astrophysics subdepartment, The Denys Wilkinson Building, 
Keble Road, Oxford OX1 3RH}

\begin{document}
\date{January 20, 2009: To be published in MNRAS}
\pagerange{\pageref{firstpage}--\pageref{lastpage}} \pubyear{2009}
\maketitle
\label{firstpage}

\begin{abstract}
The evolution of number density, size and intrinsic colour is
determined for a volume-limited sample of visually classified
early-type galaxies selected from the HST/ACS images of the GOODS
North and South fields (version 2). The sample comprises $457$
galaxies over $320$~arcmin$^2$ with stellar masses above $3\cdot
10^{10}$\Msun in the redshift range 0.4$<$z$<$1.2. Our data allow a
simultaneous study of number density, intrinsic colour distribution and
size. We find that the most massive systems ($\simgt 3\cdot
10^{11}M_\odot$) do not show any appreciable change in comoving number
density or size in our data.  Furthermore, when including the results
from 2dFGRS, we find that the number density of massive early-type
galaxies is consistent with no evolution between z=1.2 and 0,
i.e. over an epoch spanning more than half of the current age of the
Universe.  Massive galaxies show very homogeneous {\sl intrinsic}
colour distributions, featuring red cores with small scatter. The
distribution of half-light radii -- when compared to z$\sim$0 and
z$>$1 samples -- is compatible with the predictions of semi-analytic
models relating size evolution to the amount of dissipation during
major mergers. However, in a more speculative fashion, the
observations can also be interpreted as weak or even no evolution in
comoving number density {\sl and size} between 0.4$<$z$<$1.2, thus pushing major
mergers of the most massive galaxies towards lower redshifts.
\end{abstract}


\section{Introduction}
\label{sec:intro}

During the past decades the field of extragalactic astrophysics has
undergone an impressive development, from simple models that were
compared with small, relatively nearby samples to current surveys extending over
millions of Mpc$^3$ at redshifts beyond z$\sim$1 along with numerical models
that can probe cosmological volumes with the aid of large supercomputers.
However, in the same period of time, our knowledge of the 'baryon physics'
relating the dark and luminous matter components has progressed much 
slower, mainly due to the highly non-linear processes that complicate
any ab initio approach to this complex problem. 

The evolution of the most massive galaxies constitutes one of the best
constraints one can impose on the modelling of galaxy
formation. Within the current paradigm of galaxy growth in a
$\Lambda$CDM cosmology, massive galaxies evolve from subsequent
mergers of smaller structures. The most massive galaxies are
early-type in morphology and are dominated by old stellar populations,
with a tight mass-metallicity relation and abundance ratios suggesting
a quick build-up of the stellar component \citep[see e.g. ][]{ren06}.
On the other hand, semi-analytic models of galaxy formation predict a
more extended assembly history (if not star formation) from major
mergers. By carefully adjusting these models, it has been possible to
generate realizations that are compatible with the observed stellar
populations in these galaxies \cite[e.g. ][]{kav06,deluc06,bow06}

In this paper we study the redshift evolution of a sample of the most
massive early-type galaxies from the catalogue of \citet{egds09},
which were visually selected from the {\sl HST}/ACS images of the
GOODS North (HDFN) and South (CDFS) fields \citep{giav04}.  Our data
set complements recent work exploring the issue of size and stellar
mass evolution
\cite[e.g. ][]{Bun05,McIn05,fran06,fon06,Borch06,brown07,Truj07,vdk08}.  The
coverage (320~arcmin$^2$), depth ($1\sigma$ surface brightness limit
per pixel of $24.7$~AB~mag/arcsec$^2$ in the $i$ band) and
high-resolution (FWHM$\sim 0.12$~arcsec) of these images allow us to
perform a consistent analysis of the redshift evolution of the
comoving number density, size and intrinsic colour of these galaxies.

%%%%%%%%%%%%%%%
% Figure 1
%%%%%%%%%%%%%%%
\begin{figure*}
\begin{center}
\includegraphics[width=5in]{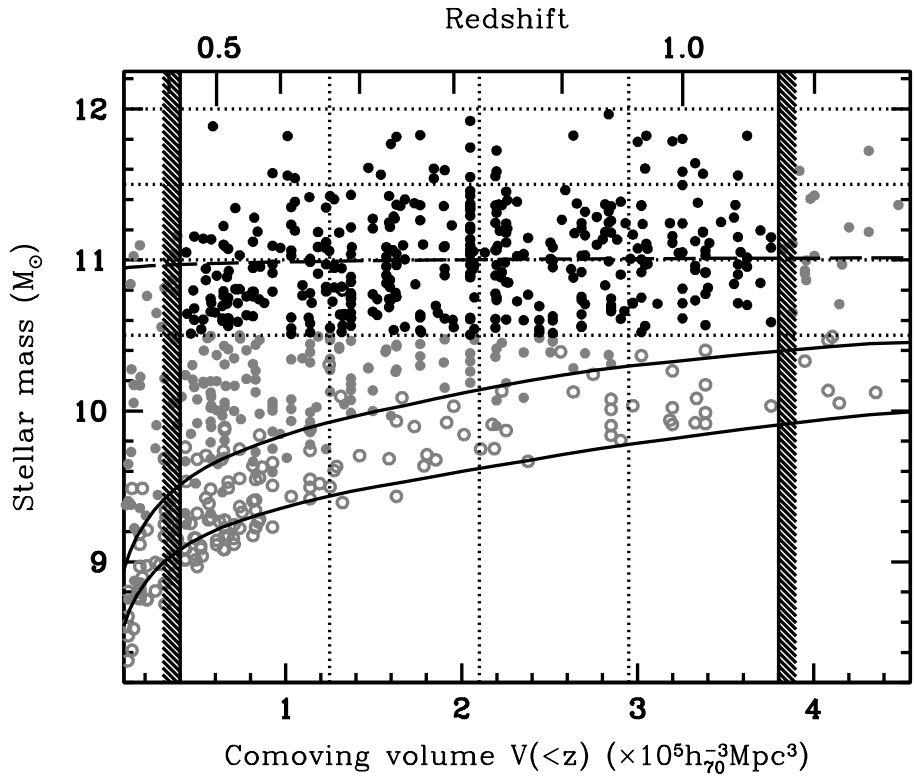}
\end{center}
\caption{Sample of massive spheroidal galaxies extracted from the v2.0
ACS/HST images of the GOODS North and South fields (Ferreras et al. in
preparation). We show in black the volume-limited sub-sample used in
this paper. We further subdivide this sample into three mass bins,
between $3\cdot 10^{10}$\Msun and $10^{12}$\Msun (stellar mass),
and in four redshift bins (top axis) centered around z=0.5, 0.7, 0.9
and 1.0 (corresponding to look-back times of 5, 6.3, 7.3, and 7.7 Gyr
before present, respectively). The dashed line tracks the
characteristic stellar mass ($M_*$) from GOODS-MUSIC \citep{fon06}. The curved
solid lines are the limiting ($i_{AB}=24$) masses for two exponentally
decaying star formation histories with solar metallicity, started at
redshift z$_F$=3 and with formation timescales $\tau=1$ (top) and
$8$~Gyr \citep[models of ][]{bc03}. Our sample is classified according to
a best fit template which roughly separates into `red' and 'blue' galaxies,
represented in this figure as filled and hollow circles, respectively.}
\label{fig:sample}
\end{figure*}
%%%%%%%%%%%%%%%

\section{The sample}
\label{sec:sample}

The {\sl HST}/ACS images of the GOODS North and South fields (v2.0)
were used to perform a visual classification of spheroidal
galaxies. This is a continuation of \citet{fer05} -- that was
restricted to the CDFS field. However, notice that our sample does
{\sl not} apply the selection based on the Kormendy relation, i.e. the
only constraint in this sample is visual classification. The analysis
of the complete sample is presented in \citet{egds09}.  Over the
$320$~arcmin$^2$ field of view of the North and South GOODS/ACS
fields, the total sample comprises $910$ galaxies down to $i_{\rm
AB}=24$ mag (of which 533/377 are in HDFN/CDFS). The available
photometric data -- both space and ground-based -- were combined with
spectroscopic or photometric redshifts in order to determine the
stellar mass content. Spectroscopic redshifts are available for 66\%
of the galaxies used in this paper. The photometric redshifts have an
estimated accuracy of $\Delta (z)/(1+z)\sim 0.002\pm 0.09$
\citep{egds09}. Stellar masses are obtained by convolving the
synthetic populations of \citet{bc03} with a grid of exponentially
decaying star formation histories \cite[see appendix B of][for
details]{egds09}.  A \citet{chab03} Initial Mass Function is
assumed. Even though the intrinsic properties of a stellar population
(i.e. its age and metallicity distribution) cannot be accurately
constrained with broadband photometry, the stellar mass content can be
reliably determined to within $0.2-0.3$~dex provided the adopted IMF
gives an accurate representation of the true initial mass function
\cite[see e.g.][]{fsb08}.

The sizes are computed using a non-parametric approach that measures
the total flux within an ellipse with semimajor axis $a_{\rm
TOT}<1.5a_{\rm Petro}$. The eccentricity of the ellipse is computed
from the second order moments of the surface brightness distribution.
The half-light radius is defined as R$_{50}\equiv\sqrt{a_{50}\times
b_{50}}$, where $a_{50}$ and $b_{50}$ are respectively the semimajor
and semiminor axes of the ellipse that engulfs 50\% of the total
flux. Those values need to be corrected for the loss of flux caused by
the use of an aperture \cite[see e.g.][]{gra05}. We used a synthetic
catalogue of galaxies with Sersic profiles and the same noise and
sampling properties as the original GOODS/ACS images to build fitting
functions for the corrections in flux and size. The corrections depend
mostly on R$_{50}$ and, to second order, on the Sersic index. Most of
this correction is related to the ratio between the size of the object
and the size of the Point Spread Function of the observations. The
dependence with Sersic index (or in general surface brightness slope)
is milder and for this correction the concentration \cite[as defined
in][]{ber00} was used as a proxy.

We compared our photometry with the GOODS-MUSIC data \citep{graz06} in
the CDFS. Our sample has 351 galaxies in common with that catalogue, and 
the difference between our
total+corrected $i$-band magnitudes and the total magnitudes from
GOODS-MUSIC is $\Delta i\equiv i_{\rm ours}- i_{\rm MUSIC}=-0.17\pm 0.16$~mag. 
This discrepancy is mostly due to our corrections of the total flux. 
A bootstrap method using synthetic images show that our corrections are
accurate with respect to the true total flux to within 0.05~mag, and to within 9\% in
half-light radius \cite[see appendix A of][]{egds09}.
Our estimates of size were also compared with the GALFIT-based
parametric approach of \citet{gems} on the GEMS survey. Out of 133
galaxies in common, the median of the difference defined as
$($R$_{50}^{\rm ours}-$R$_{50}^{\rm GEMS})/$R$_{50}^{\rm ours}$ is
$-0.01 \pm 0.16$ (the error bar is defined as the semi-interquartile
range).

%%%%%%%%%%%%%%%
% Figure 2
%%%%%%%%%%%%%%%
\begin{figure*}
\begin{center}
\includegraphics[width=5in]{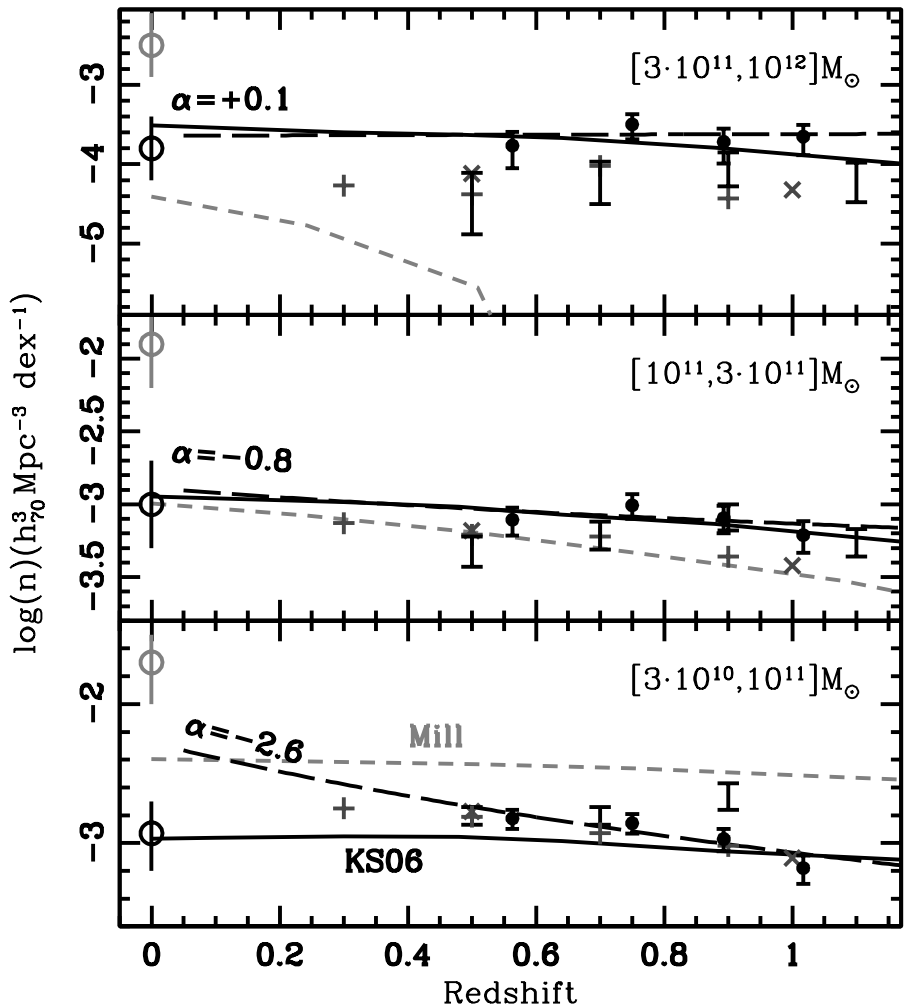}
\end{center}
\caption{ Evolution of the comoving number density of massive
early-type galaxies. Our sample (filled dots) is shown for the three
mass bins as labelled in the three panels.  The points and error bars
show the average and standard deviation for each redshift bin, adding
in quadrature Poisson noise and the uncertainty in the stellar mass
estimate. Other points from the literature are included, namely
GOODS-MUSIC \citep[$\times$ signs;][]{fon06}, Palomar/DEEP2
\citep[vertical segments;][]{con07} and COMBO17 \citep[+
signs;][]{Borch06}. The points at z=0 are estimates from 2dFGRS
restricted to (spectroscopically classified) early-type galaxies
\citep[black/grey open circles corresponding to their mean density
sample/full volume;][]{crot05}. Notice the density of the most massive
galaxies ({\sl top}) does not evolve over the whole redshift range,
spanning a look-back time of about $8$~Gyr.  The black solid lines are
model predictions from \citep{ks06a}, the grey dashed lines are from
the Millennium simulation \citep[e.g.][; no morphology
segregation]{deluc06} and the black dashed lines are fits to a power
law: $n\propto (1+z)^\alpha$, with $\alpha$ given in each panel.  }
\label{fig:logn}
\end{figure*}
%%%%%%%%%%%%%%%

We focus here on a volume-limited sample comprising early-type
galaxies with stellar mass $M_s\simgt 3\times 10^{10}$\Msun. This sample is
binned according to fixed steps in comoving volume (a standard
$\Lambda$CDM cosmology with $\Omega_m=0.3$ and $h=0.7$ is used
throughout). The complete sample of 910 galaxies from \citet{egds09} is
shown in figure~\ref{fig:sample}. Solid (open) circles represent
early-type galaxies whose colours are compatible with an older
(younger) stellar population. This simple age criterion is based on a
comparison of the observed optical and NIR colours with the
predictions from a set of templates with exponentially decaying star
formation histories, all beginning at redshift z$_{\rm F}=3$, with
solar metallicity. The ``old'' population is compatible with formation
timescales $\tau\simlt 1$~Gyr \citep[see][for details]{egds09}. The
black dots in the figure correspond to the sample of $457$ galaxies
used in this paper.

%%%%%%%%%%%%%%%
% Figure 3
%%%%%%%%%%%%%%%
\begin{figure*}
\begin{center}
\includegraphics[width=5in]{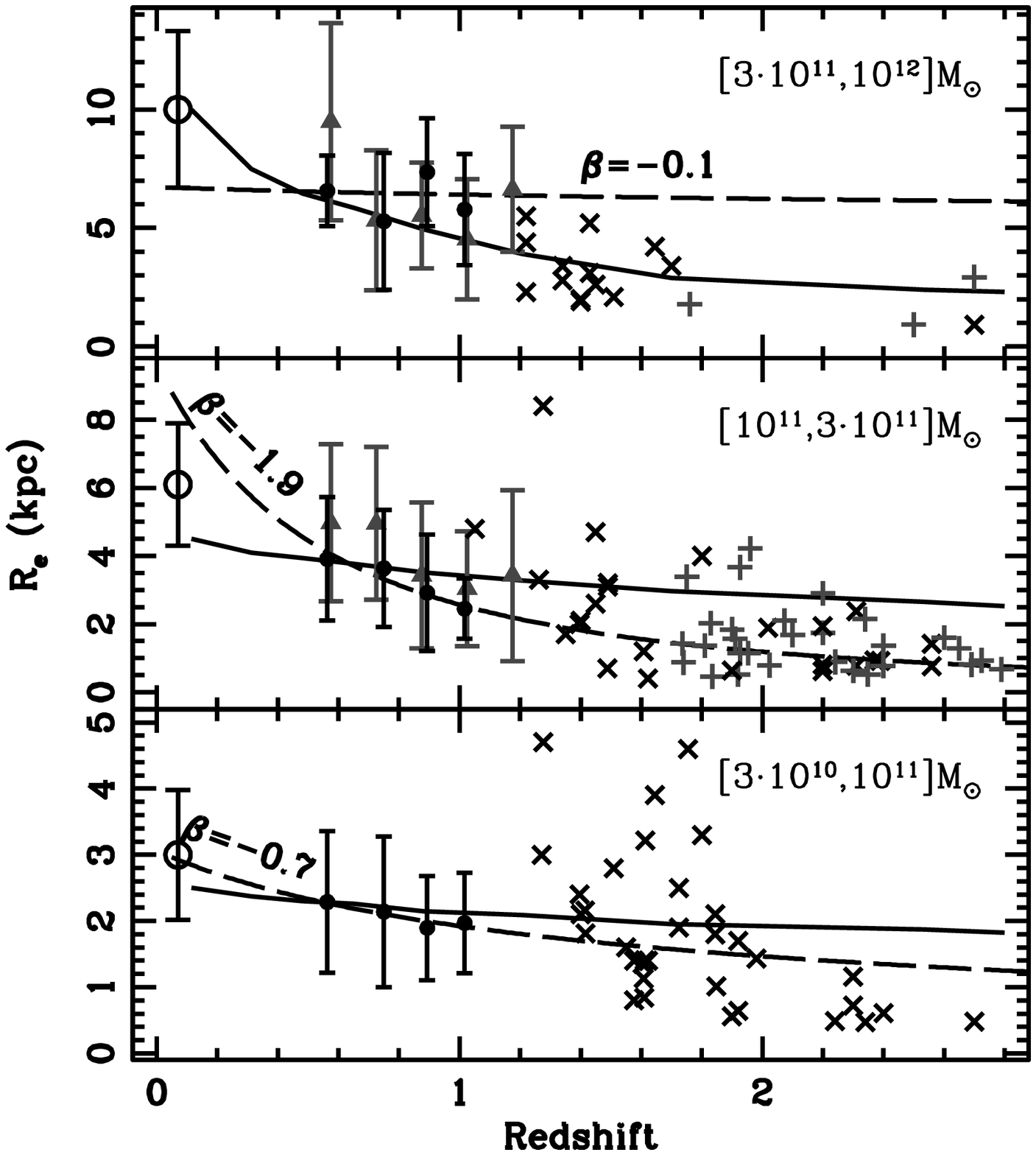}
\end{center}
\caption{
Redshift evolution of the half-light radius. Our sample is shown as
filled black circles. The data from
\citet{Truj07} is shown as grey triangles and the z$\sim$0.1 sizes of
SDSS early-type galaxies are shown as open circles \citep{Shen03}. 
At higher redshifts we include {\sl individual} measurements from
\citet[][grey `+' signs]{bui08} and from other recent work in 
the literature \citep[`x` signs:][]{zirm07,toft07,vdk08,cim08}.
The
solid lines are the predictions from \citet{ks06b}, and the dashed
lines are fits to a power law: $R_e\propto (1+z)^\beta$, with the power
law exponent given in each panel.\label{fig:Re}}
\end{figure*}
%%%%%%%%%%%%%%%

We further subdivide this sample into three mass bins, starting at
$\log ({\rm M}_s/$\Msun$)=10.5$ with a width $\Delta\log ({\rm
M}_s/$\Msun$)=0.5$~dex. For comparison, the characteristic stellar
mass from the mass function of the GOODS-MUSIC sample is shown as a
dashed line \citep{fon06}, although we warn that the GOODS-MUSIC 
masses are calculated using a \citet{salp55} IMF, which will give a systematic
0.25~dex overestimate in $\log$M$_s$ with 
respect to our choice of IMF. Our sample is safely away from the
limit imposed by the cut in apparent magnitude ($i_{AB}\leq 24$).  The curved solid
lines give that limit for two extreme star formation histories, corresponding
to the ``old'' and ``young'' populations as defined above. Notice that within our
sample of massive early-type galaxies there are NO galaxies whose colours are
compatible with young stellar populations (i.e. open circles).

\section{The evolution of massive galaxies}
\label{sec:evol}

The redshift evolution of the comoving number density is shown in
figure~\ref{fig:logn} (black dots). The ($1\sigma$) error bars include
both Poisson noise as well as the effect of a 0.3~dex uncertainty in
the stellar mass estimates. These uncertainties are computed using a
Monte Carlo run of 10,000 realizations. The figure includes data from
GOODS-MUSIC \citep{fon06}, COMBO17 \citep{Bell04} and Pal/DEEP2
\citep{con07}. At z=0 we show an estimate from the segregated 2dFGRS
luminosity function \citep{crot05}. We take their Schechter fits for
early-type galaxies within an environment with a mean density defined
by a contrast -- measured inside radius $8h^{-1}$~Mpc -- in the range
$\delta_8=-0.43\cdots+0.32$ (black open circles).  In order to
illustrate possible systematic effects in 2dFGRS, we also include the
result for their full volume sample as grey open circles.  The 2dFGRS
data are originally given as luminosity functions in the rest-frame
$b_J$. We took a range of stellar populations typical of early-type
galaxies in order to translate those luminosities into stellar
masses. The error bars shown for the 2dFGRS data represent the
uncertainty caused by this translation from light into mass over a
wide range of stellar populations (with typical M/L($b_J$) in the
range $7\cdots 12$ \Msun / \Lsun ).  The black solid lines show
semi-analytic model (SAM) predictions from \citet{ks06a}. Their SAM
follows the merging history of dark matter halos generated by the
Extended Press-Schechter formalism down to a mass resolution of
$M_{\rm min}=5 \times 10^9$ \Msun , and follows the baryonic physics
within these halos using recipes laid out in \citet[][and references
therein]{kb05}. The grey dashed lines are the predictions from the
Millennium simulation \citet[e.g. ][]{deluc06}.  This model is
extracted from their web-based 
database\footnote{\footnotesize\tt http://www.mpa-garching.mpg.de/millennium}, 
and is not segregated with respect to galaxy morphology. This explains
the excess number density in the low-mass bin (bottom panel). In the
two higher mass bins most of the galaxies have an early-type
morphology. The predictions of the Millennium simulation are in
agreement with the middle bin -- i.e. masses between $10^{11}$ and
$3\cdot 10^{11}$M$_\odot$.  However, for the most massive bin, the
sharp decrease in density with redshift of the models is in remarkable
disagreement with the observations.  In contrast, \citet{ks06a}
predict a nearly constant density at the highest mass bin out to z$<$1.

The main reason for this discrepancy is that AGN feedback in the
Millennium simulation prohibits the growth of massive galaxies by gas
cooling and subsequent star formation in order to reproduce the right
colour-bimodality and the luminosity function at z=0. As shown in
\citet{ks08} the existence of a characteristic mass scale for the
shut-off of star formation will lead to dry merging being the main
mechanism for the growth of massive galaxies. In that respect the
evolution of the number density of massive galaxies in the Millennium
simulation is mainly driven by mergers. The difference between that
model and \citet{ks06a} is probably due to the different
merger rates in their models. The Millennium simulation predicts 
a lower major merger rate compared to \citet{ks06a}
almost by a factor 10 (Hopkins et~al., in preparation). 

%%%%%%%%%%%%%%%
% Figure 4
%%%%%%%%%%%%%%%
\begin{figure*}
\begin{center}
\includegraphics[width=5in]{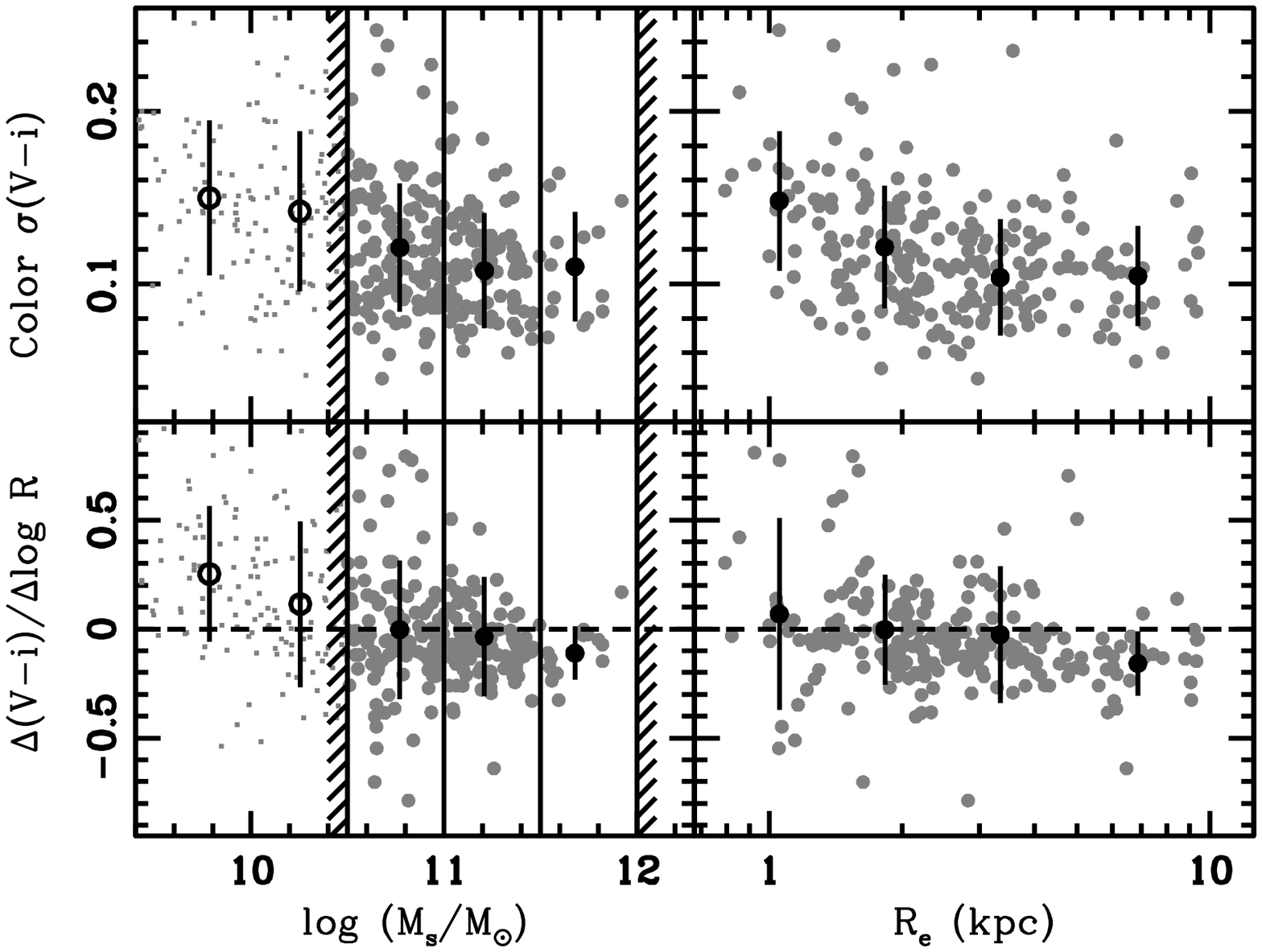}
\end{center}
\caption{The radial colour gradient ($\Delta($V$-$i$)/\Delta\log R$;
bottom) and RMS scatter (top) is shown as a function of stellar mass
({\sl left}) and projected physical half-light radius ({\sl right}).
Negative (positive) gradients imply galaxies with red (blue)
cores. The three mass bins used in this paper are illustrated by the
vertical lines and the hatched regions in the left-hand panels.  The
average and RMS values within each bin is shown by the black dots and
error bars. Hollow dots in the left-hand panels give the values for
two bins outside of our adopted mass range, for comparison. The figure
shows the trend of a reduced scatter -- concentrating the distribution
towards a slightly red core -- as mass increases.\label{fig:clr}}
\end{figure*}
%%%%%%%%%%%%%%%

Figure~\ref{fig:Re} shows the redshift evolution of the half-light
radius. Our methodology follows a non-parametric approach avoiding the
degeneracies intrinsic to profile fitting. Nevertheless, we compared
our size estimates with those using a parametric approach like GALFIT
\citep{gems} and there is good agreement (see~\S\ref{sec:sample}). Our
data (black dots) are compared with \citet[][grey triangles]{Truj07}
and with a z$\sim$0 measurement from the SDSS \citep[][taking their
early-type sample]{Shen03}. The error bars give the RMS scatter of the
size distribution within each mass and redshift bin.  The lines
correspond to the models of \citet{ks06b}. These models associate size
evolution to the amount of dissipation encountered during major
mergers along the merging history of an early-type galaxy. The points
at high redshift (z$>$1.2) correspond to {\sl individual} measurements
from the literature (see caption for details). In all the comparisons
shown in this paper with work from the literature, we have checked
that the initial mass functions used are similar, so that stellar masses are
compared consistently. All results quoted either use a
\citet{chab03} IMF or functions very close to it in terms of the total
mass expected per luminosity unit, which -- for early-type systems --
mainly reduces to the shape of the low-mass end of the IMF. Other functions
used in the quoted data were \citet{kro93,kro01} or \citet{bg03}. Only for 
the GOODS-MUSIC data \citep{fon06} the Salpeter IMF (1955) was used,
which will always give a systematic overestimate of $\sim 0.25$~dex
in stellar mass with respect to the previous choices given its
(unphysical) extrapolation of the same power law down to the low
stellar mass cutoff \cite[see e.g.][]{bc03}. A single-law Salpeter IMF is an
unlikely choice for the stellar populations in early-type galaxies
as shown by comparisons of photometry with kinematics \citep{cap06}
or with gravitational lensing \citep{fsb08}.

Similarly to the density evolution, we 
also apply a simple power law fit only to our data points: 
R$_e\propto (1+z)^\beta$. The solid lines give those best fits, and the
power law index is given in each panel. Taking into account all data
points betweeen z$=$0 and z$\sim$2.5 one sees a clear trend of
decreasing size with redshift for all three mass bins. However our data
suggest milder size evolution for the most massive early-type galaxies
between z$=$1.2 and z$=$0.4, corresponding to a 4~Gyr interval of cosmic time.

The depth and high spatial resolution of the ACS images also allow us
to probe in detail the {\sl intrinsic} colour distribution of the
galaxies (i.e. the colour distribution within each galaxy).  We follow
the approach described in \citet{fer05} which, in a nutshell,
registers the images in the two bands considered for a given colour,
degrades them by the Point Spread Function of the other passband, and
perfoms an optimal Voronoi tessellation in order to achieve a S/N per
bin around $10$ while preserving spatial resolution.  The final binned
data is used to fit a linear relation between colour and $\log
(R/R_e)$ from which we determine the slope and the scatter about the
best fit (using a biweight estimator).  Figure~\ref{fig:clr} shows the
observer-frame V$-$i colour gradient ({\sl bottom}) and scatter ({\sl
top}) as a function of stellar mass ({\sl left}) and half-light radius
({\sl right}). The black dots correspond to binned data in stellar
mass, showing the average and RMS value within each bin.  Notice the
significant trend with increasing stellar mass towards redder cores
(i.e. more negative colour gradients) and small scatter. The colour
gradient is in most cases nearly flat, and only for the lowest mass
bin do we find significantly large gradients. For comparison, we also
show as small grey dots a continuation of the original sample from
\citet{egds09} towards lower stellar masses.  Blue cores (positive
colour gradients) dominate in spheroidal galaxies below
$10^{10}$\Msun. The homogeneous intrinsic colour distribution thereby
suggests no significant star formation and a fast rearranging process
of the stellar populations if mergers take place during the observed
redshift range. Notice this sample only targets objects visually
classified as early-type galaxies.  The early phases of major merging
are therefore excluded from our sample. Nevertheless, the number
density at the massive end (upper panel of figure~\ref{fig:logn}) 
does not change significantly between z=0 and
z$\sim$1, already suggesting that major merger events must be rare
over those redshifts.

\section{Discussion and Conclusions}
\label{sec:discussion}

Using a volume-limited sample of massive spheroidal galaxies from the
v2.0 ACS/HST images of the GOODS North and South fields we have
consistently estimated the number density, size and intrinsic colour
distribution over the redshift range 0.4$<$z$<$1.2. In combination
with other samples we find a significant difference in the redshift
evolution according to stellar mass, in agreement with recent work
based on other samples or different selection criteria \citep[see
e.g.][]{Bun05,McIn05,fran06,fon06,Borch06,brown07,Truj07,vdk08}.  The
most massive galaxies -- which impose the most stringent constraints
on models of galaxy formation -- keep a constant comoving number
density between z$\sim$1 and 0 (i.e. over half of the current age of
the Universe) but present a significant size evolution, roughly a
factor 2 increase between z=1 and 0.  Note, however that within our
sample, there is no significant size evolution over the redshift range
z=0.4$\cdots$1.2. It is by extending the analysis to higher redshifts
that the size evolution shows up at the most massive bin
\citep[e.g. ][]{vdk08,bui08}. When velocity dispersion is added to the
analysis, a significant difference is found in the $\sigma$-R$_e$
distribution between z=0 and z=1, suggesting an important change in
the dynamics of these galaxies \citep{vdwel08}.

Some of the semianalytic models of massive galaxy evolution
\citep{ks06a,ks06b} are in good agreement with these
observations. These models follow the standard paradigm of early-type
galaxy growth through major mergers, with the ansatz that size
evolution is related to the amount of dissipation during major
mergers. The decreasing evolution in the comoving number density at
high masses is explained within the models by a balance between the
'sink' (loss due to mergers of massive galaxies generating more
massive galaxies) and 'source' terms (gain from mergers at lower mass)
over the redshifts considered. One could argue that the sink terms
would generate a population of extremely massive galaxies (above a few
$10^{12}$\Msun), possibly the central galaxies within massive groups
or clusters. However, this population -- with predicted comoving
number densities below $10^{-6}$Mpc$^{-3}$ -- are very hard to study
with current surveys. Furthermore, environment effects in these
systems will complicate the analysis of size
evolution \citep[e.g.][]{ko08}.

It is important to note that the lack of evolution in the number
density relates to the bright end of the luminosity function.
\citet{fab07} found a significant change in the number density of {\sl
red} galaxies with redshift. However, they also emphasize that this
change does not refer to the most luminous galaxies.  If we include
all mass bins in our sample, we do find a significant decrease in the
number density with redshift, as the lower mass bins -- which
contribute the most in numbers -- do have a rather steep decrease in
density (see figure~\ref{fig:logn}).  This difference suggests that
the (various) mechanisms playing a role in the transition from blue
cloud to red sequence must be strongly dependent on the stellar mass
of the galaxies involved.

In a more speculative fashion, our data  are
also suggestive of weak or even {\sl no evolution} in the number
density of the most massive early-type galaxies over a redshift range
0.4$<$z$<$1.2. This would imply a negligible role of major mergers at
the most massive end for z$>$0.4, thereby pushing this stage of galaxy
formation towards lower redshifts \citep{ks08}.  Another speculative
scenario for the evolution of massive spheroidal galaxies would
involve negligible major mergers at these redshifts and a significant
amount of minor mergers which will 'puff up' the galaxy. Minor mergers
are considered the cause of recent star formation observed in NUV
studies of early-type galaxies \citep{kav07}.  Larger surveys
of Luminous Red Galaxies are needed to confirm or disprove this
important issue.

%%%%%%%%%%%%%%%
% Acknowledgments
%%%%%%%%%%%%%%%

\clearpage
\onecolumn

%%%%%%%%%%%%%%%
% End of Document
%%%%%%%%%%%%%%%

\end{document}